\documentclass[nofootinbib,prl,amsmath,amssymb,twocolumn,floatfix]{revtex4-2}

\usepackage{multirow}
\usepackage{romannum}
\usepackage{color}
\usepackage{graphicx}
\usepackage{dcolumn}
\usepackage{bm}
\usepackage{siunitx}
\usepackage[normalem]{ulem}

\begin{document}

\newpage
\noindent {\large This is the accepted version of:\\
Jacopo M. De Ponti, Luca Iorio, Gregory J. Chaplain, Alberto Corigliano, Richard V. Craster, and Raffaele Ardito, \textit{Tailored Topological Edge Waves via Chiral Hierarchical Metamaterials}, Phys. Rev. Applied \textbf{19} (2023), 034079.\\
The final publication is available at: 
\textcolor{blue}{https://journals.aps.org/prapplied/abstract/10.1103/PhysRevApplied.19.034079}}

\vspace{20cm}


\title{Tailored Topological Edge Waves via Chiral Hierarchical Metamaterials}

\author{Jacopo~M.~De Ponti$^{1,*}$,  
 Luca~Iorio$^{1}$, Gregory~J.~Chaplain${^2}$, Alberto Corigliano${^1}$\\
 Richard~V.~Craster$^{3,4,5}$, Raffaele~Ardito$^{1}$
}

\affiliation{$^1$ Dept. of Civil and Environmental Engineering, Politecnico di Milano, Piazza Leonardo da Vinci, 32, 20133 Milano, Italy}
\affiliation{$^2$ Centre for Metamaterial Research and Innovation, Department of Physics and Astronomy, University of Exeter, Exeter EX4 4QL, United Kingdom}
\affiliation{$^3$ Department of Mathematics, Imperial College London, 180 Queen's Gate, South Kensington, London SW7 2AZ \\
$^4$ Department of Mechanical Engineering, Imperial College London, London SW7 2AZ, UK \\
$^5$ UMI 2004 Abraham de Moivre-CNRS, Imperial College London, London SW7 2AZ, UK}

\affiliation{$^*$Corresponding author: jacopomaria.deponti@polimi.it}


\begin{abstract}
Precise manipulation of the direction and re-direction of vibrational wave energy is a key demand in wave physics and engineering. We consider the paradigm of a finite frame-like structure and the requirement to channel energy away from critical regions, leaving them vibration-free, and redirect energy along edges towards energy concentrators for damping or energy harvesting. We design an exemplar frame metamaterial, combining two distinct areas of wave physics. Firstly, topological edge states taking an unconventional tetrachiral lattice. We control these highly localised protected edge states leveraging a hierarchy of scales through the addition of micro-resonators that impose tuneable symmetry breaking and reconfigurable mass. This allows us to achieve precise positional control in the macro-scale frame lattice, thereby opening opportunities for robust signal transport and vibration control. Experiments, theory, simulation are all utilised to provide a comprehensive analysis and interpretation of the physics.
\end{abstract}
 
 \maketitle

\begin{figure*}[t!]
    \centering
    \includegraphics[width = 1\textwidth]{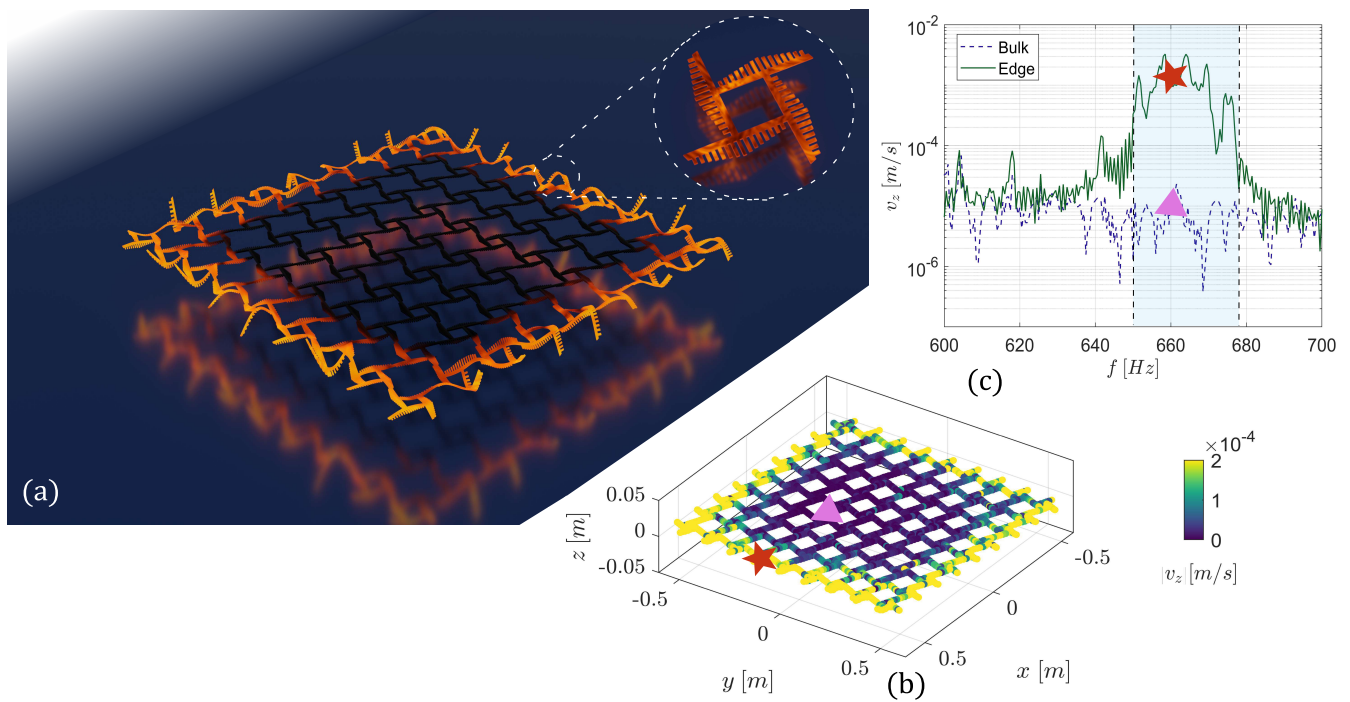}
    \caption{(a) Rendering from the numerical solution of the displacement field of the edge mode at the boundary of the lattice, together with an enlargement on the unit cell. The lattice is excited using an out-of-plane force on the edge, able to activate flexural wave propagation. (b) Experimental results obtained with a Scanning Laser Doppler Vibrometer (SLDV) on the real specimen, confirming strong agreement with numerical simulations. Experimental tests are performed by suspending the structure on a frame and providing the input using an electrodynamic shaker.  (c) Logarithmic amplitude of the experimental displacement field depending on frequency. The two curves report the bulk (blue) and the edge (green) solutions. Dashed vertical lines denotes the central part of the bulk band gap.}
    \label{Fig.1}
\end{figure*}

\section{Introduction}
The investigation of topological states of matter has attracted growing interest with multiple realizations in photonics \cite{Khanikaev2012, Lu2014, Khanikaev2015, Khanikaev2017} and phononics \cite{Mousavi2015,Huber2016}. Initiated from the field of topological insulators in quantum-mechanical systems \cite{Moore2010, Hasan2010}, protected edge (or interfacial) surface states have now percolated into the physical platforms of classical wave propagation. In the context of elasticity and  acoustics, various  topology-based mechanisms have been investigated, including  non-reciprocity  \cite{Nassar2020}, wave pumping \cite{Cheng2020}, one-way edge waves \cite{Wang2015}, and beam-splitting \cite{Makwana2018}, to name a few; an attractive property surrounding protected edge states is related to their resilience to backscattering and disorder \cite{Miniaci2018}, making them particularly important for signal transport and wave control \cite{Cha2018}. Protected bands exist within bulk band gaps and stem from broken symmetries within a periodic system; the topological nature of the Bloch bands is defined by topological invariants, most notably the Berry phase \cite{Berry1984, Xiao2010, Xiao2015, Yin2018} and the Chern number \cite{Hatsugai1993}. The nature of the symmetry breaking delineates two classes of topological insulators, that is related to the preservation of time reversal symmetry (TRS). Active topological materials \cite{Wang2015, Souslov2017, Shankar2022} have been inspired by the quantum Hall effect (QHE), where time-reversal symmetry (TRS) is broken by external fields \cite{Haldane1988}. Conversely, passive topological materials \cite{Miniaci2018,Mousavi2015,Susstrunk2015,ChaplainTopological2020} originated from the quantum spin Hall effect (QSHE) \cite{Kane2005, Bernevig2006}, where symmetry breaking is achieved by spin-orbit interactions (TRS is preserved). More recently, a number of studies have exploited valley degrees of freedom to achieve quantum valley Hall effect (QVHE) analogues \cite{Xiao2007, Jung2011, Pal2017, Vila2017, Ganti2020} in classical wave systems. QVHE relies on the breaking of space inversion symmetry (SIS), whilst maintaining TRS. This is an important difference with respect to QHE and QSHE, that permits realistic applications (that do not require complex mechanisms to achieve TRS breaking), and this is particularly appealing for vibrational wave control. 

\begin{figure*}[t!]
    \centering
    \includegraphics[width = 1\textwidth]{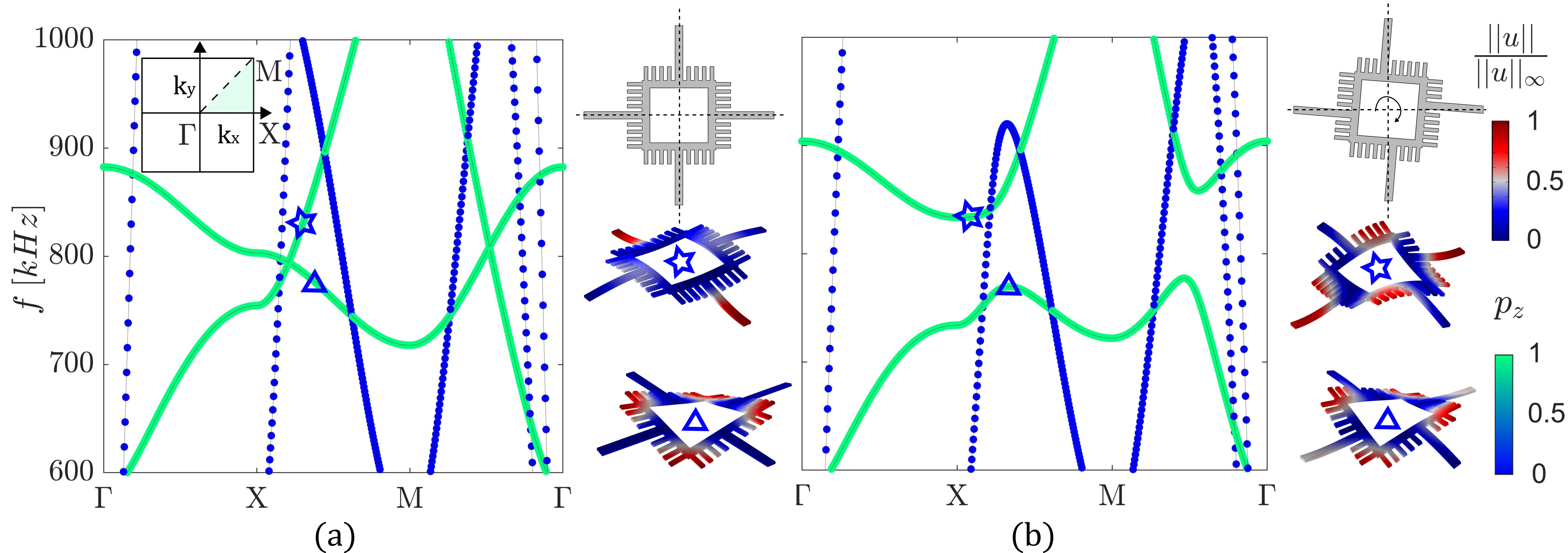}
    \caption{Vertically polarised dispersion relation for the symmetric (degenerate) (a) and the shifted (lifted degeneracy) (b) cell, with $p4m$ and $p4$ plane symmetry (wallpaper) groups respectively, computed along the Irreducible Brillouin Zone (IBZ) shown inset. The colormap of the dispersion curves shows the relative polarisation of the cell, denoted by $p_z$, with green points corresponding to vertical (out-of-plane) polarisation whilst blue refers to horizontal (in-plane) polarisation. Only the vertically polarised (green) curves are relevant, as the structure is excited to only generate out-of-plane wave propagation. The points of accidental degeneracy are lifted when the mirror symmetry is broken and the flexural waves now exhibit a bandgap (b). The geometry and eigensolutions of the cells involved are also shown, with a displacement colormap from blue to red denoting the total absolute displacement field normalised with respect to its maximum. In the symmetric configuration flexural and torsional waves are decoupled (blue star and triangle) (a), they couple when mirror symmetry is broken (b).}
    \label{Fig.2}
\end{figure*}

Within the passive topological materials framework, we tailor protected edge states in an elastic frame constructed from unit cells, see Fig. \ref{Fig.1} and Fig.  \ref{Fig.2}. 
Given the necessity of developing edge modes that propagate in the frequency region defined by the band gaps of the bulk modes, we opt for a tetrachiral structure, that allows for large band gaps in the low-frequency spectrum \cite{Bacigalupo2016, Bacigalupo2017}. We demonstrate that tetrachiral lattices do not only possess peculiar static features, but also non-trivial topological states motivated by the extension of well known topological properties of hexagonal lattices to square lattices \cite{He2014, Makwana2019, Makwana2019Optics, Makwana2020, Drost2017, Jiang2019}. In addition, the obtained edge states can be precisely tuned leveraging graded arrays of masses within each unit cell. By doing so, we introduce a hierarchy of levels allowing us to design the internal microstructure of the unit cell to obtain highly controllable edge bands.
Control over low frequency vibration is both challenging, and of practical importance. We are drawn to lightweight frame metamaterials that enable such control over a low frequency operating range, and the result is a candidate structure for addressing practical situations involving vibration isolation in ultra-precision machining \cite{Syam2018}, energy harvesting \cite{Zhao2022},  and lossless signal transport for sensing applications \cite{Xie2015}.

Separate from topological wave physics, another recent advance has been to take photonic or phononic crystals and then grade them spatially to create rainbow trapping \cite{Tsakmakidis2007,Gan2011,Chaplain2020Delineating,Ni2014,DePonti2021} or reflection devices \cite{Colombi2016,DePonti2019,Chaplain2020Tailored}. Once robust edge states are established, the combination of these with such graded structures fulfils a requirement of vibrational devices, that is to deliver maximum energy to pre-allocated positions within the lattice. Combining rainbow trapping with topological wave physics has very recently been achieved \cite{Ungureanu2021,Lu2022,Elshahat2021, ChaplainTopological2020}; we take these concepts to exemplify the advantages of the proposed design in tailoring edge waves.

\section {Design Paradigm} 

To breathe life into the concept of hierarchical metamaterial structures for robust edge wave control, we choose a tetrachiral lattice with multiple lateral resonators that act as added masses (see Supplemental Material \cite{supplemental}, Fig. S1); this choice is motivated by the broad band gaps of the structure, the square geometry for fabrication and the attraction of employing both the beneficial characteristics of chiral lattices, combined with those of a graded systems. We only utilise this structure in the low frequency range (up to 1 kHz), in which the resonance of the lateral resonators is not achieved, and so they act as added masses. The combination of the chirality with the hierarchical mass variation brings about interesting dynamical phenomena at low frequencies. The lattice possesses topological properties, as discussed in later sections, meaning that it fosters, in a specific band gap, two coincident edge modes protected from the scattering phenomena arising due to the presence of defects. The implementation of the lateral masses is key in tailoring not only the band gap that contains the desired edge modes, but also the edge modes themselves, granting the necessary design freedom needed to engineer protected edge modes with the desired dispersion characteristics.
Fig. \ref{Fig.1} shows the final structure designed and used in experiments. We illustrate both the numerical simulation with an emission colormap (Fig. \ref{Fig.1} (a)) and the experimental results (Fig. \ref{Fig.1} (b)) measured on the real specimen using a Polytec 3D Scanning Laser Doppler Vibrometer (SLDV) that scans the entire lattice. The structure shows a well-localized edge mode at around $670$ $Hz$. The experimental velocity field in the bulk and the edge is also reported in the frequency range of interest, to visualize the confined nature of such edge state (Fig. \ref{Fig.1} (c)).

To show the illustrative design process and the relationship of the observed edge states with topologically protected states, we first consider simpler, demonstrative unit cells before presenting the final structure that has been both investigated numerically and experimentally characterised, as shown in Fig. \ref{Fig.1}. 
In particular the importance of mirror symmetry, the creation of Dirac points, and the opening (or `gapping') of them by breaking mirror symmetry is illustrated in the cells shown in Fig. \ref{Fig.2}. These cells exemplify the importance of topology, and we then tune the band gap to lower frequencies by adjusting the mass and arm positions with an intermediate cell along the design path illustrated in Fig. \ref{Fig.3}. The final cell, with optimal design, is shown in Fig. \ref{Fig.4} and a ribbon geometry is used to demonstrate the edge states.

\section{Cell topology}
We design the lattice to control flexural wave propagation as they are ideal candidate waves for energy transport at low frequency, and can be easily excited.
To engineer topologically protected edge modes, it is first necessary to find and analyse a unit cell that is then periodically repeated to create an infinite lattice (the bulk medium); this is amenable to Floquet-Bloch theory and one extracts dispersion relations that relate the frequency to the Bloch wavenumber; the geometry we start with, Fig. \ref{Fig.2}, has points of degeneracy in the dispersion relation i.e. band crossings. These points are necessary for the creation of a bulk band gap, i.e. a band of frequencies for the infinite lattice in which waves do not propagate, that is generated by lifting the degeneracy through the coupling of the two modes. The cell shown in Fig. \ref{Fig.2}(a) has a dispersion relation that shows two points of so-called accidental degeneracy \cite{He2014, Makwana2019Optics} in the range $X$ - $M$ and $M$ - $\Gamma$, which denotes the high symmetry directions of the Irreducible Brillouin Zone (IBZ), as shown in the inset of Fig. \ref{Fig.2}(a). Since we are interested in the out-of-plane behaviour, we colour the dispersion curves by using the vertical polarization $p_z$, computed as the ratio between the maximum out-of-plane and in-plane displacement along the entire unit cell domain to delineate the in- and out-of- plane dominated modes.

\begin{figure*}[t!]
    \centering
    \includegraphics[width = 1\textwidth]{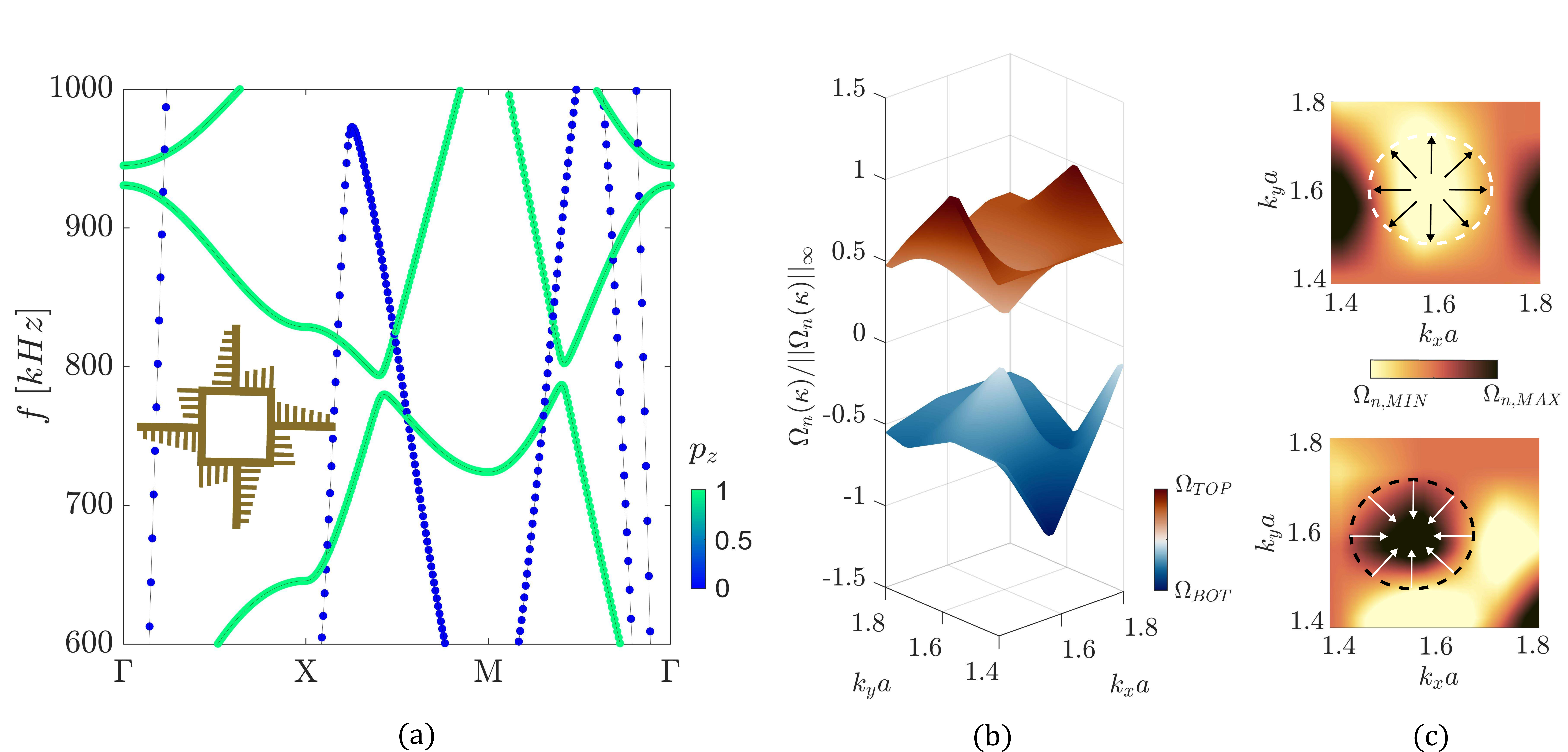}
    \caption{(a) Vertically polarised dispersion relation of the shifted cell that possesses the $p4$ symmetry group.  The cell is topologically equivalent to the degenerate case seen in Fig.~\ref{Fig.2}. The cell geometry is shown inset. (b) 3D surface view of the Berry curvatures of the top and bottom bands near the lifted degeneracy point shown in (a). The opposite sign of the cones along the bottom and top dispersion branches in the neighbourhood of the lifted degeneracy identifies a critical point. This underpins a  valley-Hall analogue which guarantees topologically protected modes within the bulk band gap. (c) Top view of the Berry curvature cones of the two modes. The gradient of the field is reported with white and black arrows, emphasising the opposite sign of the top and bottom cones.}
    \label{Fig.3}
\end{figure*}

For the out-of-plane modes that we concentrate upon, the green dispersion curves, we utilise both flexural and torsional waves, i.e. waves with rotational polarisation along the cell axes; in the perfectly symmetric case these are decoupled, but combine when a mirror symmetry breaking is introduced.
The degeneracy is marked by the intersection of the flexural and torsional waves, both marked in green in the dispersion curves of Fig. \ref{Fig.2}(a), which do not interact when the cell is four-fold rotational and mirror symmetric, i.e. a $p4m$ wallpaper group. This is seen in Fig. \ref{Fig.2} (a) where the flexural eigensolution (blue star) is not coupled with the torsional one (blue triangle) and the modes have clearly distinct symmetries.
To create the bulk band gap that hosts the protected edge mode, the degeneracy has to be lifted by coupling the two modes. To do so, the beams that connect the squared cells are shifted in an asymmetric way thereby breaking mirror symmetry, that in turn requires a rotation of the base cell to guarantee periodicity conditions; the breaking of mirror symmetry is essential in terms of opening the band gap \cite{He2014, Makwana2019Optics}. This cell, with $p4$ symmetry group, is shown in Fig. \ref{Fig.2} (b) together with the associated dispersion curves, where we notice the lifted degeneracy that now opens a band gap as we require. Contrary to the symmetric configuration in Fig. \ref{Fig.2} (a), the two eigensolutions (blue star and triangle) are now coupled, i.e. flexural and torsional waves are mixed together. This in turn eliminates the accidental degeneracy of the two modes shifting the dispersion relations, uniting them and creating the band gap of interest;  this rotated cell already has topological protection given by the mode coupling generated by removing the mirror symmetry of the cell. To determine if an edge mode, situated in the bulk band gap, is topologically protected we evaluate the Berry curvature \cite{resta2000manifestations,Semperlotti,Pal2017}.

The Berry curvature, $\Omega_n(\boldsymbol{k})$, given by (using bra-ket notation)
\begin{equation}
    \Omega_n(\boldsymbol{k}) = -\nabla_k \times \mathfrak{Im} \{ \langle \boldsymbol{u}_n (\boldsymbol{k}) \mid \nabla_k \mid \boldsymbol{u}_n (\boldsymbol{k}) \rangle \}
    \label{eq1}
\end{equation}
evaluates how, in the reciprocal space, the eigensolution rotates in the Brillouin Zone (BZ); as a 
consequence the Chern number,  $C_n$, given by
\begin{equation}
    C_n = \frac{1}{2\pi}\int_{BZ}  \Omega_n(\boldsymbol{k}) \,d^2k
    \label{eq2}
\end{equation}
counts how many times the Berry curvature has critical points over the entire zone \cite{Fosel2017}.

\begin{figure*}[t!]
    \centering
    \includegraphics[width = 1\textwidth]{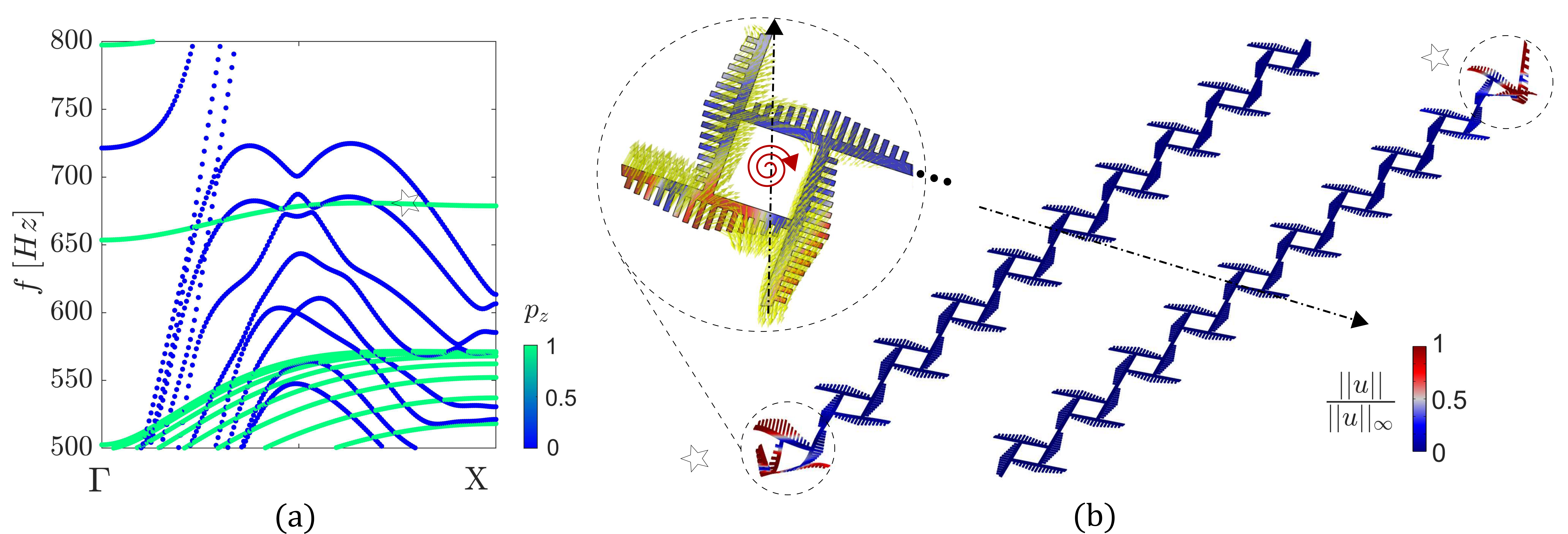}
    \caption{(a) Vertically polarised dispersion relation of the ribbon. The star indicates the degenerate edge modes and the corresponding eigensolution is reported is reported in (b). Results are obtained considering a ribbon made of 10 cells, periodically repeated along the arrow direction. An enlargement of the edge cell of the ribbon is shown with the associated magnitude of the displacement field (with a red and blue colormap). The arrow denotes the direction along which periodic boundary conditions are imposed, while the dots stands for the sequence of 9 cells in the ribbon. On top of it, the curl of the displacement field is superimposed with yellow arrows showing the chiral nature of the motion with opposite chirality between the left and right-boundary.}
    \label{Fig.4}
\end{figure*}

In these equations $\boldsymbol{k}$ is the wavevector, $\boldsymbol{u}_n (\boldsymbol{k})$ is the eigenvector associated to a certain $\boldsymbol{k}$ and to the $n$-th mode, $\boldsymbol{u}_n (\boldsymbol{k})$ is a vector containing all the normalized displacements of the nodes (from the finite element analysis) or the nodal displacement of the masses for lumped systems. The integration of the Berry curvature has to be evaluated over the complete First Brillouin Zone (FBZ), without counting twice the edges of the zone to obtain the Chern number. In certain cases a simplified calculation of the Chern number is performed near the critical points, thus giving the valley Chern number $C_v$, but generally these type of calculations are not precise and heavily dependent on the discretisation of the wavevector and mainly over strong space inversion symmetry breaking \cite{semperlotti2}. 
Here we assess the presence of the critical points of the Berry curvature, given that these points are present concurrently to the band gap only when the modes that create the band gap support topologically protected modes. To achieve the maximal band gap width, the shifts of the main beams has to be maximised, thereby increasing the coupling of the two modes; the maximum shift corresponds to the cell that we employ. Furthermore, the added lateral masses are positioned on just one side of the beams, to broaden the band gap and tune it to lower frequencies, given the strong coupling effect of mass eccentricity.  

The calculations of the Berry curvature are performed over the cell reported in the inset of Fig. \ref{Fig.3} (a). This cell still maintains a $p4$ symmetry group as the final cell of the lattice and the cones in the dispersion curves (Fig. \ref{Fig.3} (a)) are well identified thanks to the breaking of mirror symmetries. In this cell, the lateral masses are now positioned over only one side of the connecting beams, which allows us to emphasise the lifted degeneracy and to increase the band gap width. Concerning the topological protection of the edge mode, Fig. \ref{Fig.3} (b) shows the Berry curvature in the vicinity of the critical points, evaluated through Eq. (\ref{eq1}) over the opening and closing eigensolutions. To do so, the eigenvectors $\boldsymbol{u}_n (\boldsymbol{k})$ associated to the two modes are extracted from the eigensolutions over the full Brillouin Zone (BZ) of the cell in the reciprocal space. The numerical calculation is performed through ABAQUS \textsuperscript{\textregistered} finite element software, showing the two mirrored cones associated to the two modes that create the band gap. We notice that the Berry curvature in a neighbourhood of the lifted degeneracy has cones of opposite sign along the bottom and top dispersion branches. This allows us to obtain a valley Hall analogue which ensures topologically protected modes within the bulk band gap. To better visualize this result, Fig. \ref{Fig.3} (c) shows a top and bottom view of the Berry curvature, together with the gradient of the field. 

\begin{figure*}[t!]
    \centering
    \includegraphics[width = 1\textwidth]{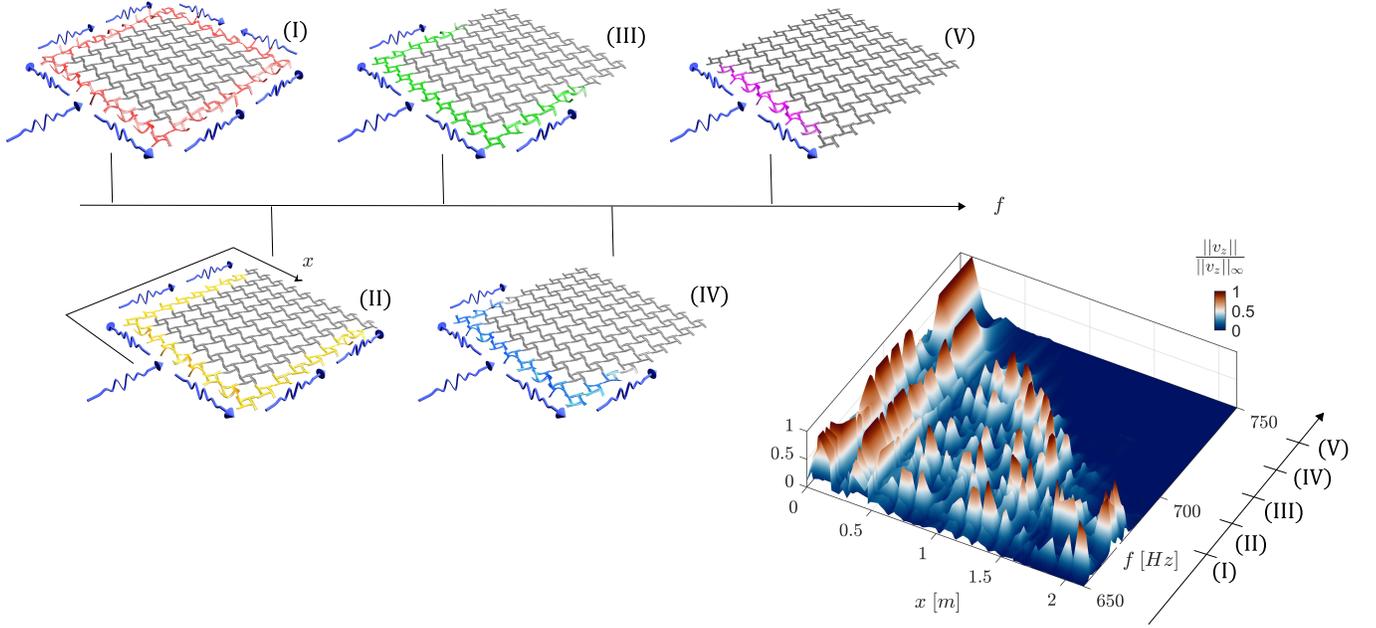}
    \caption{Results of the rainbow effect over the topologically protected edge mode. Different renderings of the numerical results are reported with a colour coating that shows where the waves, characterised by different frequency components, get blocked and confined. The blue arrows show the input wave and its propagation direction on the lattice. The numbering refers to the respective frequencies organised in ascending order. The localization and rainbow effect are shown with a corresponding time domain Fast Fourier Transform (FFT) computed over the unwrapped edge.}
    \label{Fig.5}
\end{figure*}

\section{Protected edge mode} 
Having analysed the band gap and its topological properties, we now focus on protected edge modes; these modes, existing within the bulk band gap, are a combination of torsional and flexural waves. They propagate only at the boundary of the lattice and are heavily localised on the edge cells: they decay abruptly in the bulk. Moreover, the two edge modes are identical but propagate along opposite directions. They are analysed numerically by performing a dispersion analysis on a ribbon made of one line of repeated cells. In this paper, we opt for a ribbon with 10 cells;  we note that the dispersion branch of the edge mode converges to the same result from even just five cells, but we consider 10 cells to better illustrate the strong confinement of the eigensolution. The dispersion analysis is performed considering the ribbon in Fig. \ref{Fig.4} (b), where periodic boundary conditions are applied along the direction denoted by the dashed arrow.
Fig. \ref{Fig.4} reports the results of the analysis conducted over the final, most staggered, configuration. Since the lifted eigensolution is preserved by rotation operators \cite{Lu2017}, the band gap is enlarged while preserving topological protection. The two modes are coincident, but each of them is localized only to one of the two edges. The edge mode dispersion additionally shows that the propagating mode is very slow given the near zero group velocity across most of the IBZ. This is useful for energy harvesting or sensing, since slow energy transport along the edge gives a larger interaction time with the harvesting or detection system to capture energy or signal data \cite{Chaplain2020Delineating}. Fig. \ref{Fig.4} also reports the curl of the displacement field that shows, as does the cell itself, a well defined chirality. This chiral nature of the mode supports the idea of topologically protected mode \cite{ChaplainTopological2020}, given that the wavefield is confined on the edge. The edge mode robustness is also corroborated by time-domain analyses  via implicit time
integration \cite{Hilber1977} on a defected structure (see Supplemental Material \cite{supplemental}, Fig. S2), where we notice the existence of such edge mode even in presence of large defects.  Moreover, the existence of the edge mode is confirmed by Fourier transforming time domain results (see Supplemental Material \cite{supplemental}, Fig. S4).

\section{Rainbow edge mode}
Having analysed the edge mode properties and behaviour, we now focus towards modifying its characteristics, namely its group and phase velocities and the position of the zero group velocity point within the Brillouin Zone. The aim is to tailor these properties without modifying the total added mass given by the lateral resonators, so that the band gap that fosters the protected mode is not altered between the different cells. In the cases analysed here, we define the functions for the change in length of the masses in such a way that the added mass is maintained constant, no matter the spatial distribution. In this way, the location of the edge mode remains confined in the protected bulk band gap, without significant frequency shift; thus, topological protection is preserved. The control over the dispersion properties of the edge mode is necessary to engineer its behaviour at the frequency range of interest. The change of the dispersion branch is then used to progressively modify the propagating wave characteristics, for instance creating grading effects, on the edge of the lattice (see Supplemental Material \cite{supplemental}, Fig. S3). We then demonstrate the advantages of the proposed design concept creating, as a meaningful example,  the rainbow effect. The rainbow effect engineered in this work has been constructed through the use of five different cells with the same geometry, except from the grading of masses. Specifically, we adopted the following linear grading of lengths: $l_1 \in [13,\:5]\:mm$, $l_2 \in [10.7,\:6.5]\:mm$ (inverse grading), $l_3 \in [8.7,\:8.7]\:mm$ (periodic), $l_4 \in [7.2,\:10.4]\:mm$, $l_5 \in [5,\:13]\:mm$ (direct grading). The configurations have been chosen carefully following the set of rules defined above, and maintaining the notion of an adiabatic change of the homogenised properties of the lattice, i.e. considering a slow variation of the fundamental Bloch mode of each cell along space \cite{Schnitzer2017, Johnson2002}. The edge rainbow structure has also been assembled following the idea of having two symmetric directions of propagation of the input edge wave, meaning that the overall lattice has a mirror symmetry group. The lattice mirror plane is positioned exactly at the center of one of the edges, where we apply the input wave.

Within the lattice, in the bulk, we keep unchanged the spatial distribution of the masses. This choice does not disrupt the propagating edge mode because the bulk band gap supports the edge mode even when the lateral masses are modified, highlighting that the spatial distribution can be defined and implemented by just considering the edge mode dispersion branch. Fig. \ref{Fig.5}  reports the results obtained from a FEM analysis on the described structure. It shows how a broadband input generated at the center of one edge propagates at the boundary of the structure, being confined at different spatial positions depending on its frequency. Fig. \ref{Fig.5} also highlights the localization of frequency components with respect to spatial positions over the unwrapped edge, that is a peculiar property of the rainbow effect \cite{Tsakmakidis2007}.

\noindent\textit{Conclusions}| We have shown, both numerically and experimentally, the physics of a hierarchical tetrachiral lattice that fosters topologically protected edge modes. The edge mode was engineered to obtain full control over its propagation both in space and in time. Moreover, we also obtained, as a meaningful example, the topological rainbow effect by modifying the dispersion properties of the edge mode along the boundary of the lattice, granting the localization of different frequency components along the edge. This control over the dispersion properties was obtained thanks to micro elements added to the cell (lateral masses) without changing the shape or the added mass. Future developments will assess the behaviour of the structure at higher frequencies, i.e. when the resonators begin to work dynamically at resonance.






\noindent\textit{Acknowledgements}| A.C., J.M.D.P, L.I. and R.A. acknowledge the  financial support of the H2020 FET-proactive Metamaterial Enabled Vibration Energy Harvesting (MetaVEH) project under Grant Agreement No. 952039 in supporting the research activity and the prototype realization. R.V.C. acknowledge the financial support of the FET-Open scheme BOHEME - Bio-Inspired Hierarchical MetaMaterials project under Grant Agreement No 863179. G.J.C gratefully acknowledges financial support from the Royal Commission for the Exhibition of 1851 in the form of a Research Fellowship. The authors wish to thank D. Cavazzi for the prototype realization and Prof. F. Braghin for the experimental tests.




\end{document}


\title{Tailored Topological Edge Waves via Chiral Hierarchical Metamaterials: Supplementary Information
}

\author{Jacopo~M.~De Ponti$^{1,*}$,  
 Luca~Iorio$^{1}$, Gregory~J.~Chaplain${^2}$, Alberto Corigliano${^1}$\\
 Richard~V.~Craster$^{3,4,5}$, Raffaele~Ardito$^{1}$
}

\affiliation{$^1$ Dept. of Civil and Environmental Engineering, Politecnico di Milano, Piazza Leonardo da Vinci, 32, 20133 Milano, Italy}
\affiliation{$^2$ Centre for Metamaterial Research and Innovation, Department of Physics and Astronomy, University of Exeter, Exeter EX4 4QL, United Kingdom}
\affiliation{$^3$ Department of Mathematics, Imperial College London, 180 Queen's Gate, South Kensington, London SW7 2AZ \\
$^4$ Department of Mechanical Engineering, Imperial College London, London SW7 2AZ, UK \\
$^5$ UMI 2004 Abraham de Moivre-CNRS, Imperial College London, London SW7 2AZ, UK}
\affiliation{$^*$ Corresponding author: jacopomaria.deponti@polimi.it}

\maketitle

\beginsupplement

To aid insight into the underlying physics of the protected edge mode in the hierarchical metamaterial, we provide a detailed description of the unit cell geometry, numerical models/additional analyses, sample manufacturing and experimental methods.

\section{Unit cell geometry}
\begin{figure}[h!]
    \centering
    \includegraphics[width = 0.5\textwidth]{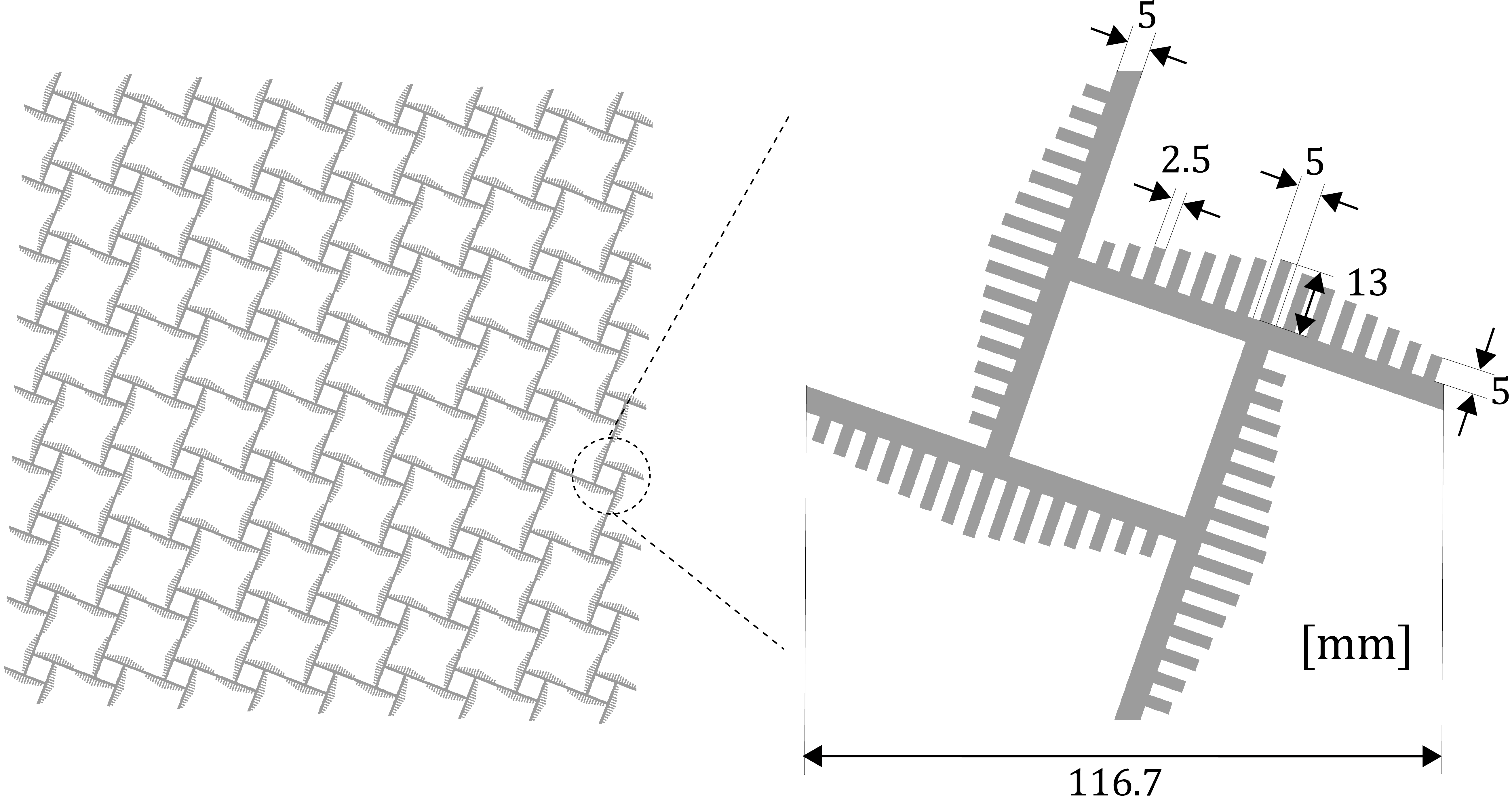}
    \caption{Finite lattice structure and unit cell geometry. The lattice is made of  9 $\times$ 9 unit cells each with a tetrachiral geometry. Within the unit cell, a set of graded masses, the short elastic beams of varying length, is introduced to enable precise control over the group velocity of the protected edge mode.}
    \label{FigS1}
\end{figure}

The finite lattice we use in modelling and experiments is obtained by periodically repeating 9 $\times$ 9 unit cells, each with dimension $a_1=116.5$ $mm$ (Fig. \ref{FigS1}). The unit cell has a tetrachiral geometry and is constructed from elastic beams $5\,mm$ wide with thickness $1.5$ $mm$.
Within the unit cell, we introduce a set of graded masses to control the group velocity of the protected edge mode; the graded masses are short beams of varying length, $2.5$ $mm$ wide, equally spaced by a distance $a_2=5\,mm$. We adopt as the grading parameter the beam length, which linearly ranges from $5$ $mm$ to $13$ $mm$, resulting in $\approx$ $13^{\circ}$ slope angle.

\begin{figure*}[t!]
    \centering
    \includegraphics[width = 1\textwidth]{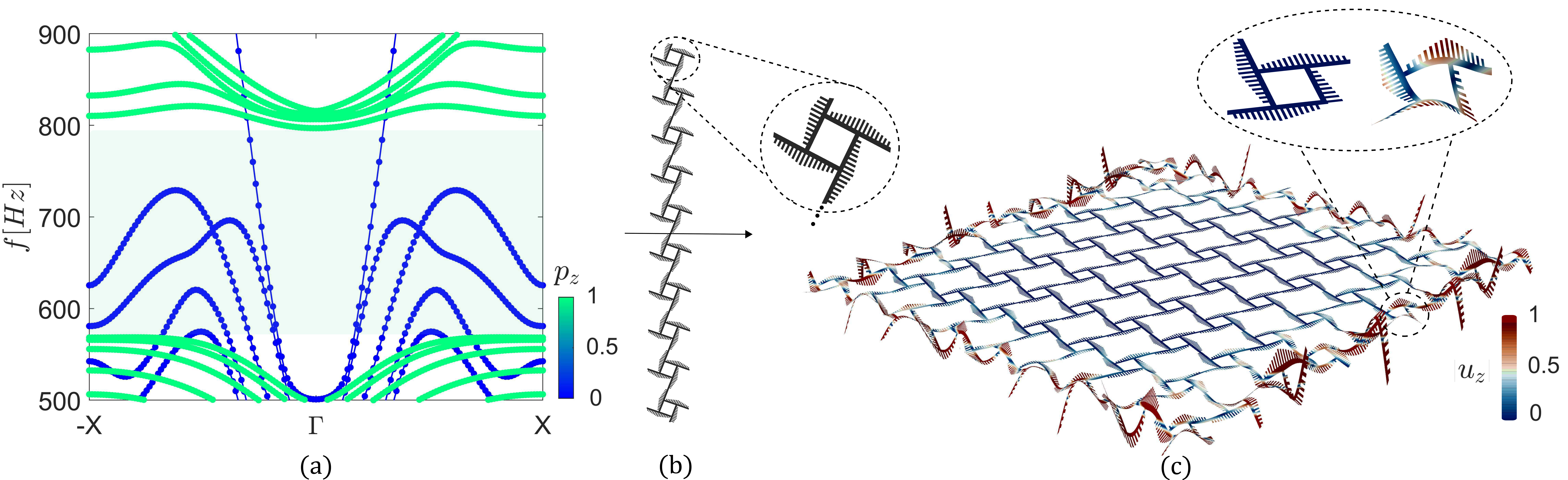}
    \caption{Corroboration of edge mode robustness by introducing a defect along the edge of the periodic lattice. (a) Dispersion relation obtained from a ribbon made of 10 cells, with the introduction of a defect at the outer arms. By cutting half of the outer arms, the edge mode (vertically polarised) within the bulk band gap disappears. The colormap of the dispersion curves shows the relative polarisation of the cell with green points corresponding to vertical (out-of-plane) polarisation, whilst blue refers to horizontal (in-plane) polarisation. (b) Corresponding ribbon structure used to compute the dispersion relation in (a). Periodic boundary conditions are imposed along the arrow direction, while the defect is introduced by removing half of the beams at the edges. (c) Wave propagation analyses along the finite lattice with a larger defect (the entire outer arms have been removed, as shown in the zoomed image). The existence of the edge mode in the defected structure corroborates the claim from theory of topological protection.}
    \label{FigS2}
\end{figure*}

\section{Edge mode robustness}
We investigate edge mode robustness by taking the unit cell of Fig. \ref{FigS1} and then modifying it by introducing a defect to it. 
\\
We eliminate, in the outermost cells of the ribbon, half the outer arm; the defect is only applied
to the edge of the outermost cells as we want to study the effect of introducing a defect on the protected edge mode.
Fig.\,\ref{FigS2}(a) shows the polarised dispersion curves for a ribbon made of 10 cells, that is periodically repeated in the direction of the arrow shown in Fig.\,\ref{FigS2}(b). In this ribbon, the  defect, shown as the insert to Fig.\,\ref{FigS2}(b), has been introduced at both edges, i.e. to the cells at the top and bottom of the ribbon, as both ends of the ribbon, when periodically extended and without a defect, support the edge mode. The defect applied on one edge only, would just suppress the edge mode on that specific edge.

The polarised dispersion in Fig.\,\ref{FigS2}(a) shows that the introduction of the defect completely destroys the edge state within the bulk band gap: a lattice created from only the periodic repetition of this ribbon structure would not support edge states. 
This study on the defect is carried out to assess whether the edge mode exists or not when the edge cells are damaged.  The dispersion results of  Fig.\,\ref{FigS2}(a) show that the edge mode cannot propagate along an infinite edge created solely from the defected cells, or at least the wave would be strongly scattered by the defects. For the protected edge modes where we only introduce isolated defects along the edge, the scattering due to the defects does not affect the propagating edge wave: the topological nature of the wave grants that the wave is impervious to scattering. For this reason, we want to assess whether the edge mode shows the robustness towards scattering of defects, confirming the numerical calculation in the main text about its topological nature. We insert a single defective cell placed at the edge in our original perfect lattice of Fig. \ref{FigS1} and explore the implications of altering the size of the defect by shortening the outer arm. 
\\
Shortening the outer arm yet further generates an ever more extreme defect and to demonstrate the extent to which we achieve robust energy transport we show the wave propagation where we have completely removed one outer arm at the lattice boundary, see Fig. \ref{FigS2} (c).

Through time domain simulations we assess that, in the presence of the defect described above (complete removal of the outer arm at one position), the edge mode still exists in the periodic lattice (see Fig. \ref{FigS2}(c)). Moreover we do not see any significant scattering effect given by this large defect.
\\
This corroborates the protected nature of the edge mode, as theoretically demonstrated in the main manuscript by means of the Berry curvature, and suggests that, in practical terms, this design is highly robust to manufacturing defects or damage.

\section{Effect of the graded microstructure}
The unit cell is designed to couple flexural and torsional waves leveraging broken symmetry.

\begin{figure}[t!]
    \centering
    \includegraphics[width = 0.48\textwidth]{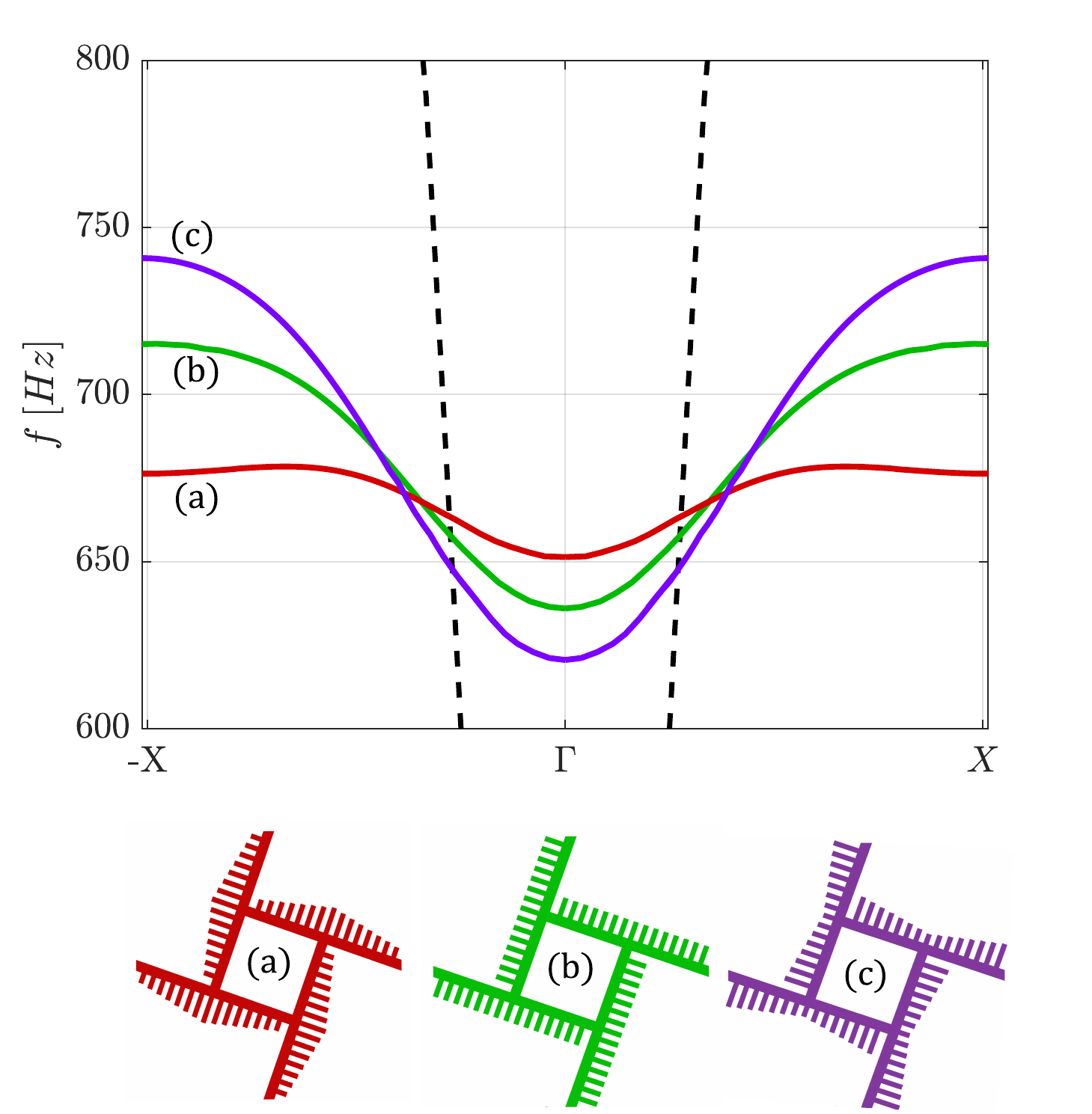}
    \caption{Tunability and grading of the masses. Dispersion curves of the edge mode on a ribbon made of 10 unit cells at varying grading of masses. By introducing different distributions of masses within the unit cell (keeping constant the total added mass) the group velocity of the edge waves is tuned and controlled. This allows to obtain, at parity of introduced mass, strong control over the wave velocity, providing a powerful tool to tailor topological edge waves.}
    \label{FigS3}
\end{figure}

Starting from a symmetric configuration (as shown in Fig. 2(b) in the main manuscript), we then offset the external beams and rotate the unit cell to guarantee periodicity. Moreover, we additionally lift the degeneracy by moving the added masses from the internal portion of the cell to the exterior beams, increasing the asymmetry. By doing so, we enlarge the band gap width, gaining a stronger confinement of the edge mode. The addition of masses allows us to control the bulk band gap both in terms of frequency and width. By adopting different gradings of masses, and keeping constant the total added mass, we can control the group velocity of the edge mode. Fig. \ref{FigS3} shows the edge mode dispersion for different distributions of masses within the unit cell. We notice that by changing the distribution of masses, we can strongly modify the group velocity of the edge mode. This concept allows us to tailor the edge mode, keeping constant the total mass of the structure. Moreover, by introducing cells with different distribution of masses along the edges of the same lattice structure, we can achieve the rainbow effect, i.e. spatial signal separation depending on frequency. Indeed, creating a sequence of cells with different mass distributions (e.g. Fig. \ref{FigS3} (a)-(c)) we open band gaps at different positions depending on the frequency content of the wave. It is important to notice that the distribution of masses has no impact on the bulk band gap. This means that, in order to tailor the group velocity of this edge mode, we just need to modify the mass distribution along the edges, without changing the internal part of the lattice.

\section{Additional numerical results}
\begin{figure}[h!]
    \centering
    \includegraphics[width = 0.5\textwidth]{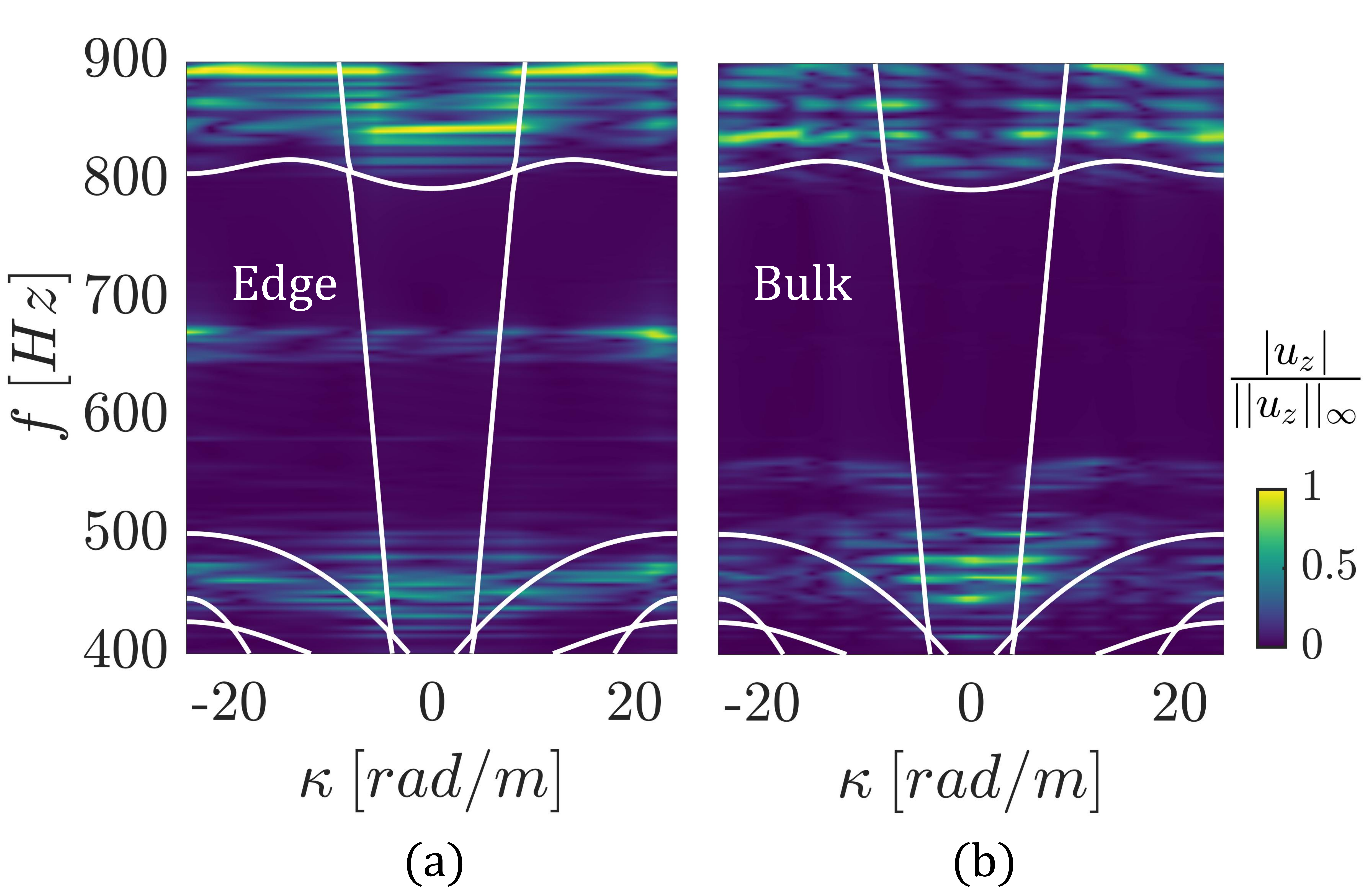}
    \caption{Two-dimensional (2D) Fast Fourier Transform from the displacement field on the edge (a) and in the bulk (b) of the lattice structure. An edge mode clearly appears on the boundary of the lattice, while disappears in the bulk. White lines represent the bulk dispersion curves from dispersion analysis.}
    \label{FigS4}
\end{figure}

To corroborate the existence of the edge mode, we complement the dispersion  analyses by the two-dimensional (2D) spatiotemporal Fast Fourier Transform (FFT) of the wavefield from time domain analyses performed in ABAQUS\textsuperscript{\textregistered}. We adopt a Finite Element discretisation using 8-node doubly curved thick shell elements with reduced integration (S8R), with 6 degrees of freedom (dof) for each node. The
analysis is performed via implicit time integration based on the Hilber-Hughes-Taylor operator \cite{Implicit}, an extension of the
Newmark $\beta$-method with a constant time increment dt = $0.05$ $ms$. The system is forced through an imposed force on the edge of the structure. The input signal is a finite burst $F(t) = F_0w(n)\sin(2\pi f_c)$ with amplitude  $F_0= 1$ $N$, Hann window $w(n)$, central frequency $f_c= 670$ $Hz$ and time duration $0.1$ $s$. We interpolate the wavefield along straight lines in order to have a proper spatial resolution of the signal. Fig. \ref{FigS4} compares the numerical FFT for a set of points along the edge (Fig. \ref{FigS4} (a)) and in the bulk of the lattice (Fig. \ref{FigS3} (b)). We notice that an edge mode is clearly identified at approximately $670$ $Hz$ on the edge, while it disappears in the bulk.

\section{Experimental setup}
The metalattice is studied experimentally by means of a Polytec 3D Scanner Laser Doppler Vibrometer (SLDV), which is able to separate the out-of-plane velocity field in both space and time. 

\begin{figure}[h!]
    \centering
    \includegraphics[width = 0.45\textwidth]{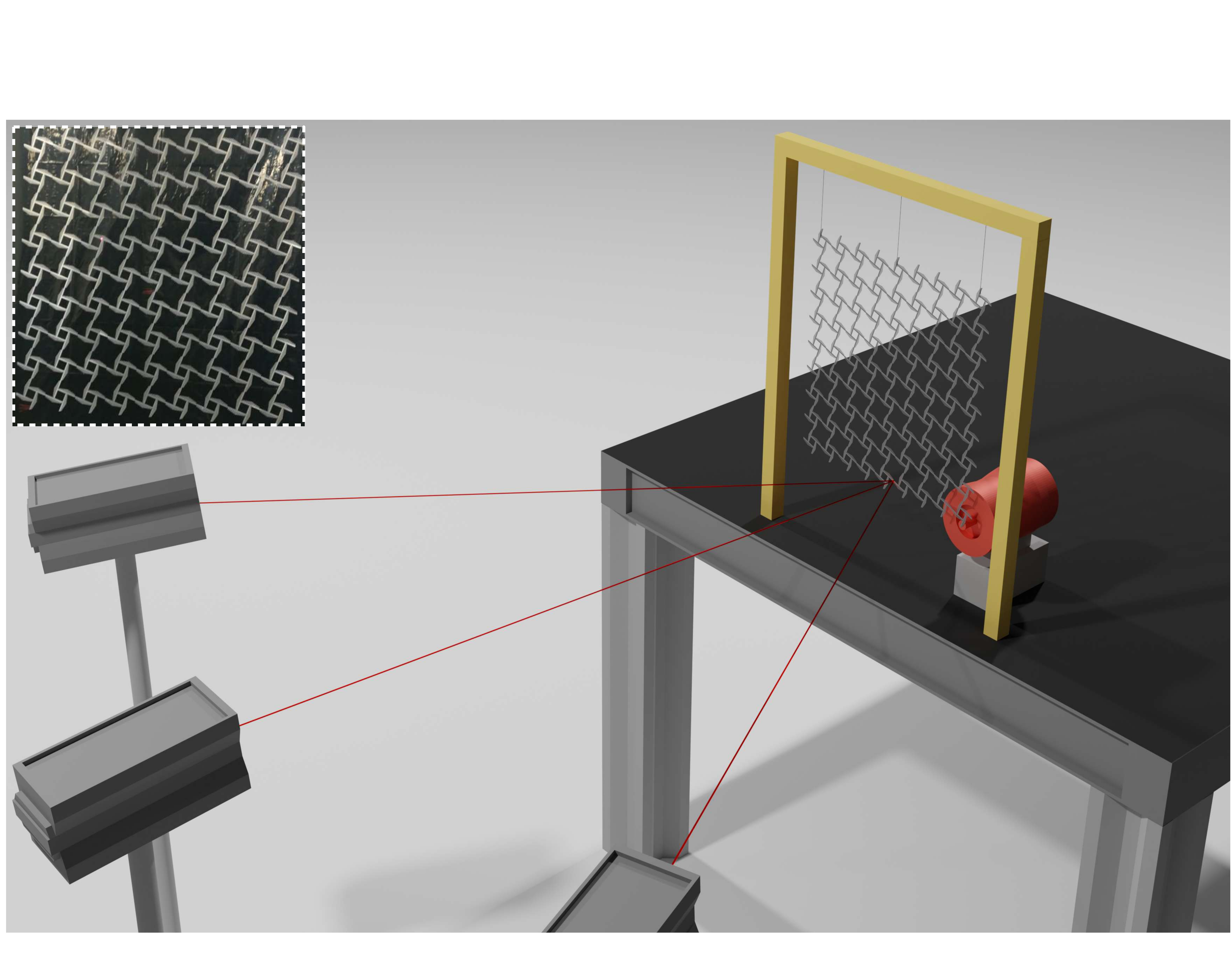}
    \caption{Experimental setup. The wavefield is measured using a 3D Scanner Laser Doppler Vibrometer (SLDV). The lattice, suspended on a frame structure, is forced using an electrodynamic shaker which is able to create an out-of-plane flexural wave. The inset on the top of the figure shows a photo of the real prototype.}
    \label{FigS5}
\end{figure}

The specimen is obtained by water-jet cutting an aluminum plate of thickness $1.5$ $mm$. The experimental setup is shown in Fig. \ref{FigS5}. The lattice is suspended on a frame structure by means of elastic cables that do not affect the dynamics of the system. The input wave is imposed by a LDS v406 electrodynamic shaker, which is attached to the bottom right corner of the structure. The input signal is thus a flexural wave applied to one arm of the free ended lattice. We perform both time domain and frequency domain analyses. In  time domain we use a narrowband source $V(t) = V_0w(n)\sin(2\pi f_c)$ with amplitude  $V_0= 2$ ${\rm V}$, Hann window $w(n)$, central frequency $f_c= 670$ $Hz$ and time duration $0.1$ $s$. For frequency domain analyses we Fourier transform time domain results, using as input a sweep signal in the range $400-900$ $Hz$ and time duration $2$ $s$; the results of such analyses are reported in Fig. 1(b) and (c) of the main manuscript.